\documentclass{aa}
\usepackage{graphicx}
\usepackage{amssymb}
\usepackage{latexsym}
\usepackage{epsfig}
\usepackage{graphics}

\begin{document}

\titlerunning{Integral field near infrared spectroscopy of two blue dwarf galaxies}
\authorrunning{Cresci et al.}

\title{Integral-field near-infrared spectroscopy of two blue dwarf galaxies: NGC~5253 and He~2-10\thanks{Based on observations obtained at the ESO-VLT under programs 074.A-9011 and 075.B-0648}}

\author{G. Cresci \inst {1,2},
          L. Vanzi \inst{3},
	  M. Sauvage \inst{4},
          G. Santangelo \inst{5,6,7,1}, \and
          P. van der Werf \inst{8}
          }

\offprints{G. Cresci}
 \institute{Osservatorio astrofisico di Arcetri, Largo E. Fermi 5, 50125 Firenze, Italy \\
 \email{gcresci@arcetri.astro.it}
 \and 
  Max-Planck-Institut fuer Extraterrestrische Physik, Postfach 1312, 85741, Germany \\
 \and
 Department of Electrical Engineering, Pontificia Universidad Catolica de Chile, Av. Vicu\~{n}a Mackenna 4860, Santiago, Chile \\  
 \and
 Laboratoire AIM, CEA/Irfu, CNRS/INSU - Universit\'e Paris Diderot, CEA-Saclay, 91191 Gif sur Yvette Cedex, France \\ 
 \and
 European Southern Observatory, Karl Schwarzschild str.2, D-85748 Garching bei Muenchen, Germany \\
 \and
 Universit\'a di Bologna, Dipartimento di Astronomia, via Ranzani 1, 40127 Bologna, Italy \\ 
 \and
 INAF - Istituto di Radioastronomia, via Gobetti 101, 40129 Bologna, Italy\\ 
 \and
 Leiden Observatory, Leiden University, PO Box 9513, 2300 RA Leiden, The Netherlands\\
}

   \date{Received ... ; accepted .... }

   \abstract{We present integral field spectroscopy in the near infrared (NIR) of He~2-10 and NGC 5253, two well known nearby dwarf irregular galaxies showing high star-formation rates. Our data provide an unprecedented detailed view of the interstellar medium and star formation in these galaxies, allowing us to obtain spatially resolved information from the NIR emission and absorption line tracers. We study the spatial distribution and kinematics of different components of the interstellar medium (ISM) mostly through the Bracket series lines, the molecular hydrogen spectrum, [FeII] emission, and CO absorptions. Although the ISM is mostly photo-excited, as derived by the [FeII]/Br$\gamma$ and $H_2$ line ratios, some regions corresponding to non-thermal radio sources show a [FeII]/Br$\gamma$ excess due to a significant contribution of SN driven shocks. In He~2-10 we find that the molecular gas clouds, as traced by CO(2-1) and $H_2$ infrared line, show consistent morphologies and velocities when studied with the two different tracers. Moreover, there is a clear association with the youngest super star clusters as traced by the ionized gas. In the same galaxy we observe a cavity depleted of gas, which is surrounded by some of the most active regions of star formation, that we interpret as a signature of feedback-induced star formation from older episodes of star formation. Finally, we measured high turbulence in the ISM of both galaxies, $\sigma \sim 30-80$ km/s, driven by the high star-formation activity. 

}

   \maketitle

   \section{Introduction}

Detailed observations of  intense star formation and mass assembly 
are a key ingredient in probing the role of various mechanisms in shaping the galaxies and moderating their evolution. Starburst galaxies are complex systems, where gas is converted into stars on a relatively short timescale with great efficiency: the theoretical progress in this area is still partly dependent on highly simplified recipes for the physical mechanisms that drive the formation of new stars and their interplay with the interstellar medium (ISM).  Building a complete theoretical picture is important to fully understand the mechanism responsible for the formation and evolution of star-forming galaxies at high redshift ($z\gtrsim1$), where the bulk of star formation in the Universe took place (e.g. Rudnick et al. \cite{rudnick96}). Important progress has been made in recent years by direct observations of distant galaxies, but at those distances even the largest sources are just a few arcsec in size (e.g. Cresci et al. \cite{cresci06}), preventing us from distinguishing the different mechanisms at work. Although perfect analogs of high-z galaxies are not available locally owing to the different conditions in the ISM, gas content, accretion rate, and chemical properties, studying star-forming galaxies at a much higher physical resolution is critical to investigate some key properties that are virtually inaccessible at high redshift. 
Nearby starburst systems have already helped in the past decades to understand some of the processes that regulate and trigger star formation on galactic scales, e.g. the interaction with close companions or large molecular clouds, and internal mechanisms such as feedback from previous generations of massive stars. Blue compact dwarf galaxies are particularly suited for this task, as they show typically sub-solar abundances, thus providing the opportunity to observe star formation in a relatively pristine environment and with less contamination by the underlying evolved stellar population than in larger galaxies. 

The population of young super massive stellar clusters (SSCs) that accounts for a large fraction of the star formation of the entire galaxy was identified in starburst systems both locally and at high-z (e.g. de Grijs et al. \cite{degrijs03}, Elmegreen et al. \cite{elmegreen05}). These clusters could be the progenitors of the old globular clusters observed in most galaxies today (de Grijs \& Parmentier \cite{degrijs07}), thus providing further evidence for the link between the starburst phenomenon and the formation of galaxies.
Moreover, star-forming regions with properties consistent with the more massive HII regions detected locally are observed up to $z\sim5$ (Swinbank et al. \cite{swinbank09}), and local dwarf galaxies have been identified as scaled down analogs of clumpy star-forming galaxies at high redshift (Elmegreen et al. \cite{elmegreen09}). Dwarf galaxies indeed show high gas fraction, high turbulence, a large Jeans length compared to their size and a clumpy appearance, similar to what is observed in high-z forming disks (see e.g. Genzel et al. \cite{genzel06},\cite{genzel08}; Cresci et al. \cite{cresci09}, Forster-Schreiber et al. \cite{natascha09}). In this picture, the clumpy morphology comes from gravitational instabilities in gas with high turbulent speed compared to the rotational one, and high gas mass fraction compared to stars. Although the mechanism providing the gas in the local and high-z galaxies might be different (mergers and interactions are required locally, while cosmological accretion through cold flows might be dominant at high redshift, see e.g. Dekel et al. \cite{dekel09}), the physical process regulating the star formation in clusters are probably the same in the two environments. 

In the effort of understanding the physical processes at play in galactic-scale star formation, we thus undertook a survey campaign of nearby dwarf starburst galaxies using integral field spectroscopy in the near infrared (NIR). This technique provides an unprecedented detailed view of the interstellar medium and star formation in these galaxies, allowing us to obtain spatially resolved information from the emission and absorption line tracers in the NIR. The results derived from the study of the nearby starburst galaxy II~Zw~40 were presented in Vanzi et al. (\cite{vanzi08}), while in this paper we discuss the observations obtained on He~2-10 and NGC~5253, two very well known examples of nearby starbursts.
Both galaxies are dwarf-irregular, with estimated distances of 9 Mpc for He~2-10 (Vacca \& Conti \cite{vacca92}) and 3.3 Mpc for NGC~5253 (Gibson et al. \cite{gibson00}): due to their relatively close distances and remarkable properties they have been studied in detail in the past.

\subsection{He~2-10}

\begin{figure*}
\centering
\includegraphics[angle=0,width=0.8\textwidth]{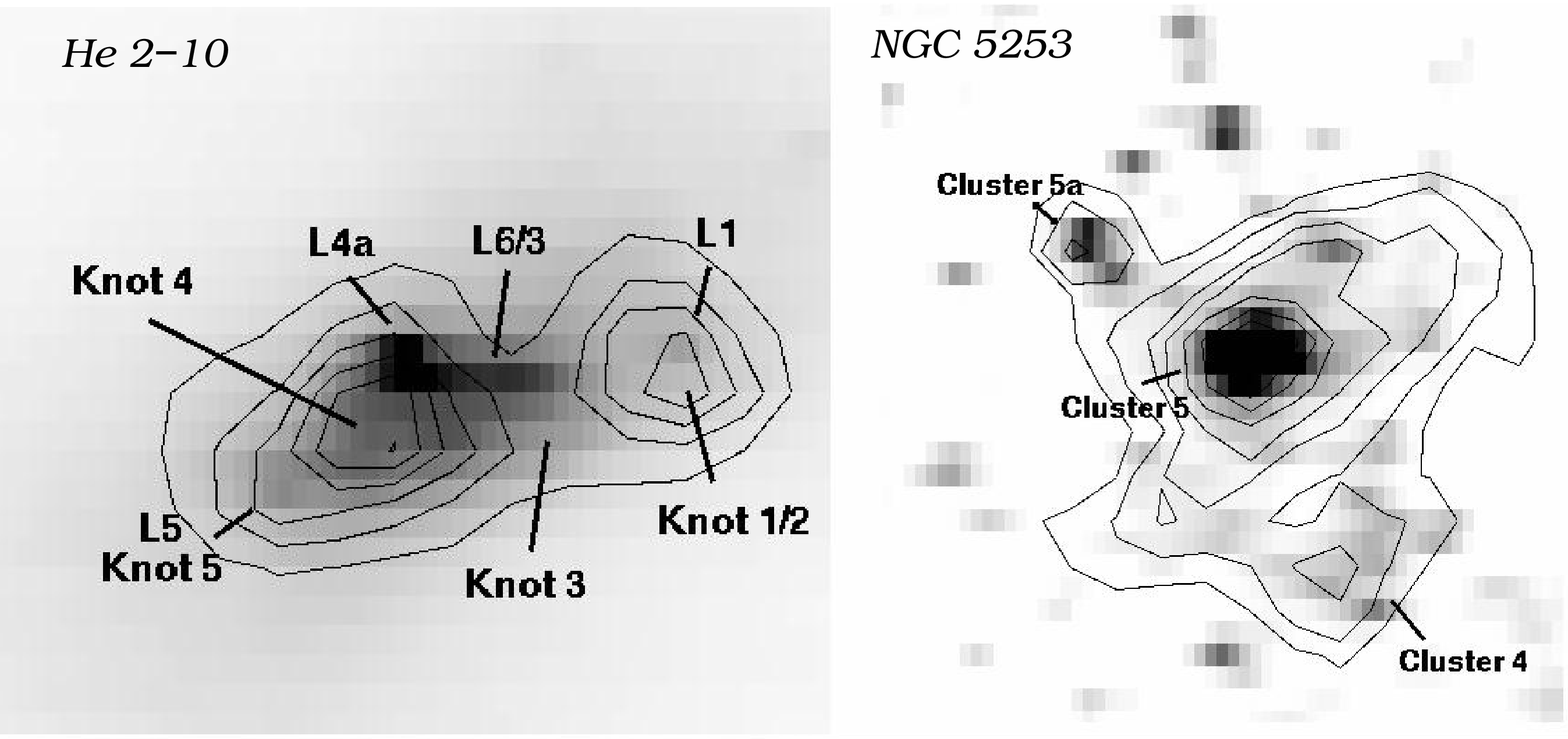}
\caption{K-band continuum images of He 2-10 (left panel) and NGC 5253 (right panel), extracted from the SINFONI datacube. The Br$\gamma$ contours are over-plotted as reference. The main clusters and star-forming regions in the field of view are identified with the naming convention from literature: for He~2-10, the clusters (L1-6) from Cabanac et al. (\cite{cabanac05}) and the radio knots (knot 1-5) from Kobulnicky \& Johnson (\cite{kob99b}) are marked, while for NGC~5253 the clusters from Calzetti et al. (\cite{calzetti97}) and Vanzi \& Sauvage (\cite{vanzi04}).}
\label{namingfig}
\end{figure*}

He~2-10 has nearly solar metalicity ($12+log(O/H)=8.93$; Kobulnicky et al. \cite{kob99a}) and hosts an intense episode of star formation as revealed by classical indicators such as the strong optical and infrared emission lines (Vacca \& Conti \cite{vacca92}, Vanzi \& Rieke \cite{vanzi97}), the Wolf-Rayet features (Vacca \& Conti \cite{vacca92}), intense mid-IR (Sauvage et al. \cite{sauvage97}) and far-IR continuum (Johansson \cite{johansson87}). Some compact radio sources were observed by Johnson \& Kobulnicky (\cite{johnson03}), who characterized their spectra as thermal, measured their sizes as $2-4$ pc and their masses in excess of $10^5\ M_{\odot}$. A number of young super massive clusters in the star-forming region of the galaxy, with typical masses up to $10^5\ M_{\odot}$ and ages of few Myr have been found in optical and UV studies (e.g. Johnson et al. \cite{johnson00}, Vacca \& Conti \cite{vacca92}). However, these observations reveal clusters that are relatively unobscured and are probably not the youngest star-forming regions. Cabanac et al. (\cite{cabanac05}, hereafter C05) observed in the IR, to penetrate the obscuring material, and detected several compact sources that could be associated with the radio sources previously mentioned, finding that most of them are ultra-compact HII regions and two are possible SN remnants (see Fig. \ref{namingfig}). 

Kobulnicky et al. (\cite{kob95}) derived a total molecular gas mass of $1.6 \cdot 10^8\ M_{\odot}$ and an atomic gas mass of $1.9 \cdot 10^8\ M_{\odot}$ through single-dish observations of CO(1-0) and CO(2-1) and VLA HI 21cm. Thanks to the interferometric resolution, they detected a CO plume extended about $30\arcsec$ to the southeast, and a similar but more extended feature in HI. They interpreted these features as a sign of a merger with a companion galaxy. This interpretation is supported by the detection of an extended component of CO(3-2) molecular gas to the northeast of the main body of the galaxy by Vanzi et al. (\cite{vanzi09}),  interpreted as additional evidence of a detached cloud expelled in the merger that He 2-10 is experiencing. 

\subsection{NGC~5253}

NGC~5253 is an irregular dwarf galaxy in the Centaurus Group and another interesting target to study young starbursts. It is indeed one of the closest starburst galaxies, with a heliocentric distance of only 3.3 Mpc (Gibson et al. \cite{gibson00}), so that $1\arcsec$ corresponds to 16 parsec, and one of the youngest starburst galaxies known (Rieke et al. \cite{rieke88}), as implied by the detection of spectral features arising from Wolf-Rayet stars (Schaerer et al. \cite{schaerer97}) and from its almost entirely thermal radio spectrum with very little synchrotron emission from supernova remnants (Beck et al. \cite{beck96}). Its metalicity is sub-solar and about $Z_{\odot}/6$ (Kobulnicky et al. 1999). 

At centimetric wavelengths, Turner et al. (\cite{turner98}) and Turner et al. (\cite{turner00}) have observed several nebulae, which they derive to be ionized by $200-1000$ O stars, suggesting that very large clusters are the preferred mode of star formation in the central regions of the galaxy. Calzetti et al. (\cite{calzetti97}) found that they span a range of masses and ages, and identified six main clusters, one of which powers a giant HII region that is the most prominent object in the galaxy (Vanzi \& Sauvage \cite{vanzi04}). Cresci et al. (\cite{cresci05}) detected more than 300 clusters combining high-resolution NIR and optical imaging, with derived ages spanning a range between 2 and 50 Myr and masses of up to $10^6\ M_{\odot}$. \\

The paper is organized as follows: Sect. \ref{obsdata} provides the details of the observations, in Sect. \ref{analysis} we describe in detail the results, in Sect. \ref{cavity} we discuss a possible origin of the cavity detected in the ISM of He~2-10, while in Sect. \ref{fine} we summarize our results.

\section{Observations and data reduction} \label{obsdata}

The observations were obtained with the integral field spectrograph SINFONI at the ESO-VLT (Eisenhauer et al \cite{frank2003}, Bonnet et al. \cite{bonnet}) as part of the ESO program 075.B-0648 (
He2-10 and NGC5253) and 074.A-9011 (
He2-10).

\begin{figure*}
\centering
\includegraphics[angle=0,width=0.9\textwidth]{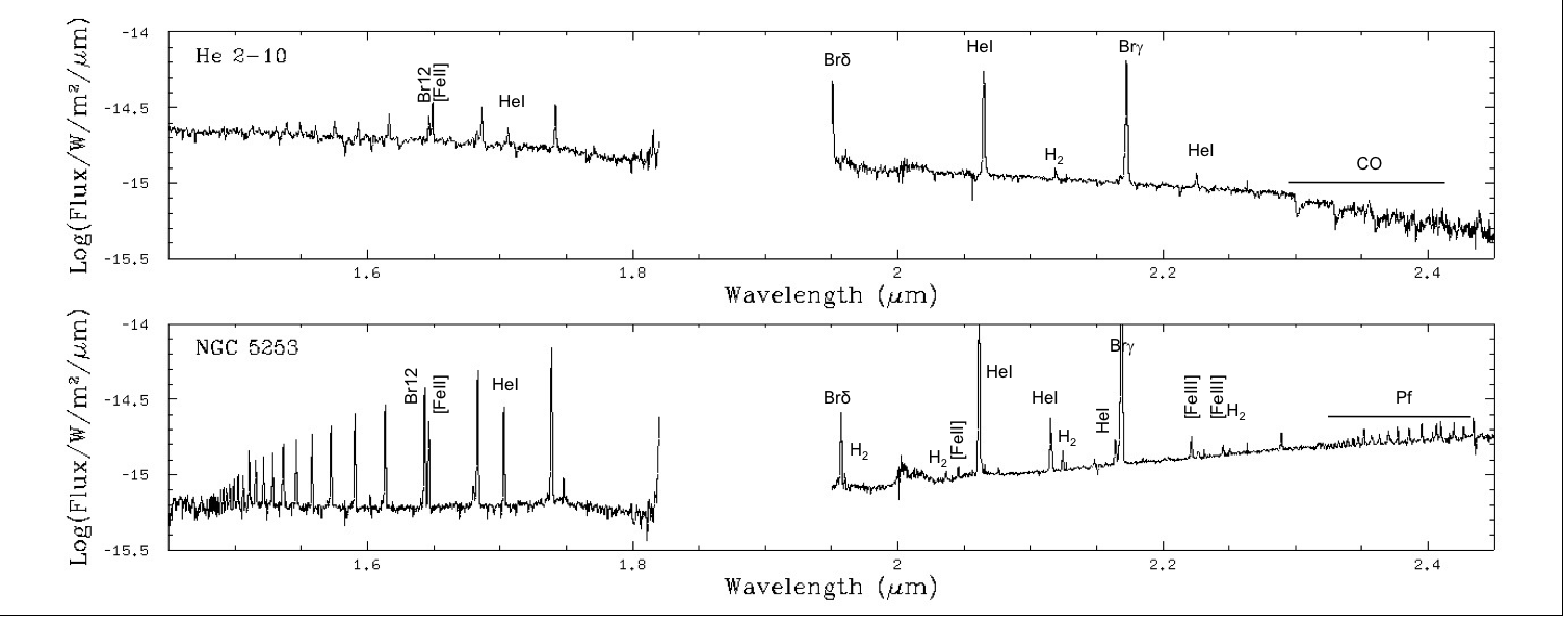}
\caption{Near-infrared H and K spectrum extracted with a circular aperture of $0.375\arcsec$ radius around the Br$\gamma$ peak of the two galaxies, with the main detected emission lines explicitly indicated, execept for the large number of lines from the Brackett series detected in the H-band. }
\label{1dspectra}
\end{figure*}

We used the $0.250\arcsec \times 0.125\arcsec$ pixel scale, which provides a field of view of $8\arcsec \times 8\arcsec$. Due to the large apparent sizes of these two galaxies, we were able to cover only part of the optical extent of the targets with SINFONI. For He~2-10  our pointing is centered on the bright central nucleus, generally referred to as region A in the literature, which is an arc of UV-bright super star clusters that dominate the emission of the galaxy (Vacca \& Conti \cite{vacca92}). In NGC~5253 our observations were centered at the brightest HII region, i.e. cluster 5 according to the notation of Calzetti et al. (\cite{calzetti97}). 

For each galaxy we obtained two data cubes using the 
H and K gratings, providing one spectral resolution R=
3000 and 4000 respectively. He2-10 was observed on 2005 March 24 and 2005 April 3 for a total integration time on-source of 20' in both H and K band. 
NGC5253 was observed on  2005 April 3, with an integration time of 10' in 
both bands.

The data were reduced with the SINFONI custom reduction package SPRED (Abuter et al. \cite{abuter}), which includes all of the typical reduction steps applied to near-IR spectra with the additional routines necessary to reconstruct the data cube. After background subtraction, which was performed with off sky frames obtained interspersed with the on-source exposures, the data were flat-fielded and corrected for dead/hot pixels. Telluric correction and flux calibration were carried out using A- and B-type stars, after removing the stellar features. Residuals from the OH line emission were minimized with the methods outlined in Davies (\cite{ric07}). 

A crude estimate of the seeing limited spatial resolution achieved was obtained with the standard star observations. These show a FWHM $\sim0.6\arcsec$ for both galaxies, corresponding to a linear resolution of $26.2$ and $9.6$ pc at the distances assumed for He~2-10 and NGC~5253 respectively. 

\begin{figure*}
\centering
\includegraphics[angle=0,width=1\textwidth]{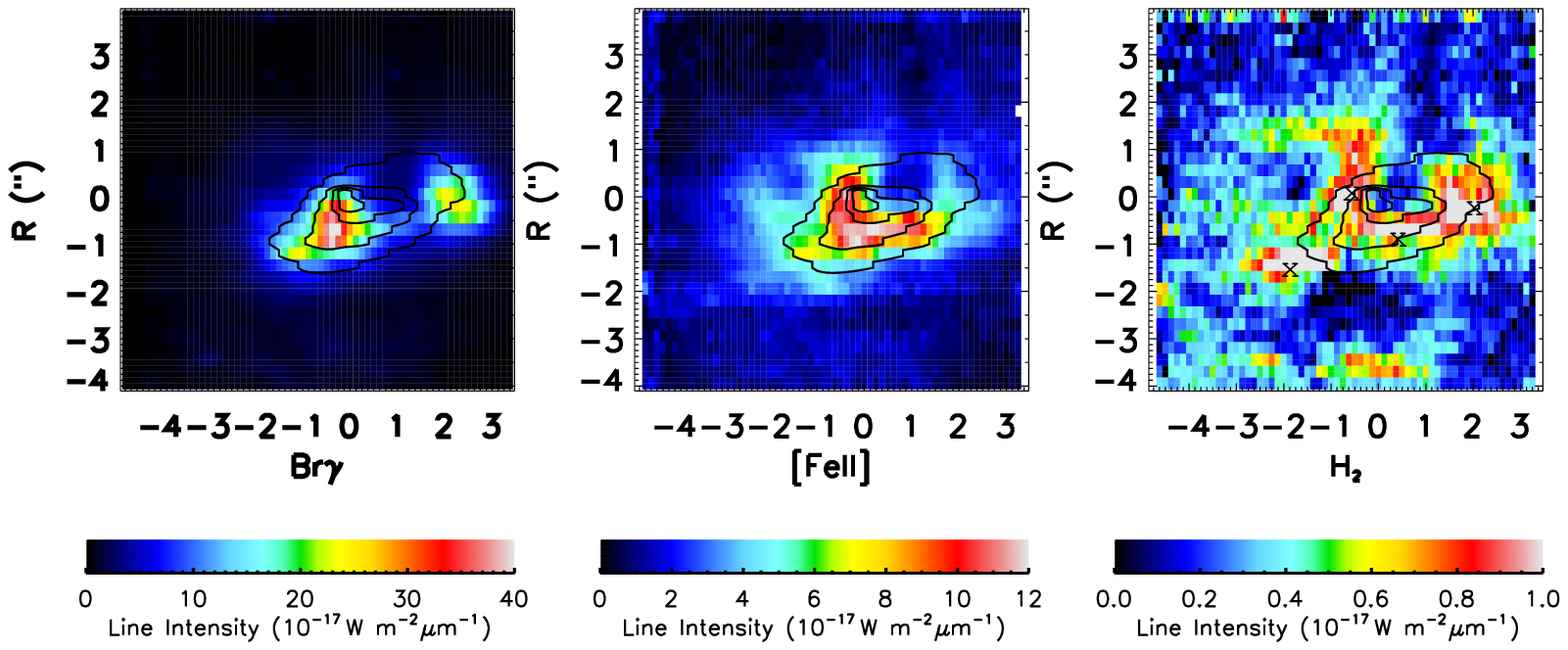}
\caption{Br$\gamma$, [FeII] and $H_2$ images of He~2-10, the contours of the K-band continuum are overplotted on each image. The coordinates are centered on the continuum peak of the brightest cluster. The crosses in the $H_2$ images mark the regions discussed in Sect. \ref{molH}}
\label{flux_he2-10}
\end{figure*}
\begin{figure*}
\centering
\includegraphics[angle=0,width=1\textwidth]{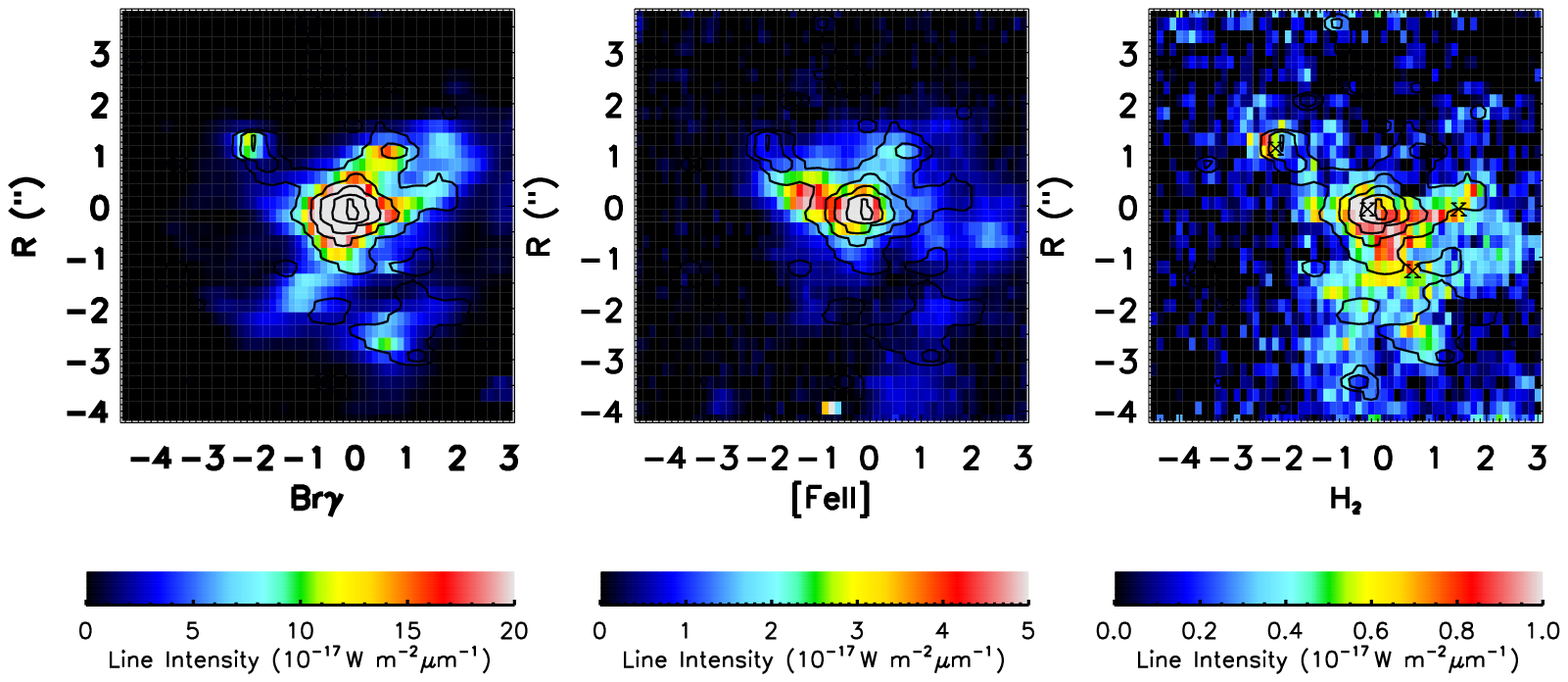}
\caption{Br$\gamma$, [FeII] and $H_2$ images of NGC~5253, the contours of the K-band continuum are over plotted on each image. The coordinates are centered on the continuum peak of the brightest cluster. The crosses in the $H_2$ images mark the regions discussed in section \ref{molH}}
\label{flux_ngc5253}
\end{figure*}
An example of the obtained very rich spectra for the two galaxies is shown in Fig.~\ref{1dspectra}. Our analysis is focused on the main visible features, mainly the emission lines of atomic hydrogen, the emission lines of molecular hydrogen, the forbidden emission lines of [FeII], and the stellar absorption features of CO. We extracted spectral images centered on the main features detected, fitting each emission or absorption line at each spaxel with an unresolved line profile (e.g. a sky line) convolved with a Gaussian, as well as a linear function to the line-free continuum. This fitting process gives the intrinsic velocity dispersion (i.e., instrumental broadening has been accounted for), the absolute velocity and the total flux in the line at that location, and these values are used in the subsequent analysis, figures, and tables. The uncertainty of the fit is estimated with Monte Carlo techniques, by refitting the best-fit Gaussian with added noise of the same statistics as the data. This is done 100 times, and the standard deviations of the best-fit parameters are used as the uncertainties (see Cresci et al. \cite{cresci09}, Forster Schreiber et al. \cite{natascha09}).

\section{Analysis} \label{analysis}

In Figs. \ref{flux_he2-10} and \ref{flux_ngc5253} we show the line maps, extracted as explained in Sect. \ref{obsdata}, in Br$\gamma$ 2.167 $\mu$m, [FeII] 1.643 $\mu$m, $H_{2}$ 2.121 $\mu$m rest wavelength for both galaxies. The contours of the K-band continuum from the SINFONI data-cubes are drawn for comparison. 
Below we will mainly use these tracers to study the properties and dynamics of the different phases of the ISM in the two galaxies: ionized gas through recombination lines of H and other species like Fe, warm molecular gas through $H_2$, and dust extinction using ratios of well-modeled lines. 

\subsection{Ionized hydrogen} \label{ionH}

The ionized gas distribution, which we can best trace in our data through the Br$\gamma$ emission, allows us to identify the main regions of star formation in the two galaxies. We refer to the first panel of Figs. \ref{flux_he2-10} and \ref{flux_ngc5253} for He~2-10 and NGC~5253 respectively.

In He~2-10 three main regions are prominent in Br$\gamma$. The central region, where the emission peaked, is also bright in H$\alpha$, which coincides with the region called L4 by C05. This area is bright and extended in the radio continuum and is identified as knot 4 by Kobulnicky \& Johnson (\cite{kob99b}, hereafter KJ99).
A second bright Br$\gamma$ region is seen to the southeast of the previous one. It is not bright in H$\alpha$, but it is detected in the I, K, and N band continuum (C05), probably due to higher extinction (see Sect. \ref{estinzione}). It coincides with source L5 of C05 and with the radio knot 5 of KJ99.
Finally, a region to the northwest of the central one is also detected in H$\alpha$ as a diffuse nebula, bright in L and 10 $\mu$m (knot L1 in C05) and coincident with the radio sources knot 1 and 2 of KJ99 (see Fig.~\ref{namingfig}).

The K continuum detected in SINFONI and shown in Figs.  \ref{namingfig}  and \ref{flux_he2-10} in contours is identical to the K image obtained by C05. The comparison of the Br$\gamma$ morphology and the continuum shows a clear difference, with the maximum of the K continuum located $0.7\arcsec$ to the NW of the Br$\gamma$ one, and a plume of three clusters bright in continuum but undetected in Br$\gamma$. These bright K sources at the center of the galaxy (L4a, L3a,b and L6 according to C05) are actually older clusters of more evolved stars (see Sect. \ref{coabsorption}).

NGC 5253 offers a significantly different view. It was already noted by Vanzi \& Sauvage (\cite{vanzi04}) and Cresci et al. (\cite{cresci05}) that the radiation emitted by this galaxy is dominated at all wavelengths by one single compact source, the cluster identified with number 5 by Calzetti et al. (\cite{calzetti97}) at the center of our pointing. This source is a super massive cluster with an age of just a few Myr and a mass of $\sim 1.2 \times 10^6 M_{\odot}$, which also dominates the emission in our Br$\gamma$ map (see Figs.~\ref{namingfig} and \ref{flux_ngc5253}). The Br$\gamma$ source is coincident with the K continuum pick. We also detect the source identified as 5a by Vanzi \& Sauvage (\cite{vanzi04}) to the NE of source 5 and cluster 4 of Calzetti et al. (\cite{calzetti97}). All the other main clusters fall outside our field of view. Yet a number of fainter sources are detected, mostly in our continuum image without a counterpart in the ionized gas map, and an extended diffuse component in the emission line.

\subsection{Ionized iron}  \label{iron}

The forbidden lines of [FeII] are emitted in partially ionized regions in thin shells at the edge of HII regions, or in more extended regions produced by shocks (Mouri et al. \cite{mouri00}). Therefore these lines have been used as tracers of young supernova remnants (SNR) in starburst galaxies and AGN (Moorwood \& Oliva \cite{moorwood88}, Oliva et al. \cite{oliva90}, Colina \cite{colina93}, Vanzi \& Rieke \cite{vanzi97}, Alonso-Herrero et al. \cite{almudina03}). However, the observation of these systems with high enough angular resolution, as well as the observation of star-forming regions in the Galaxy, allowed to establish that it is not rare to detect purely photo-excited lines in objects still dominated by young massive stars (e.g. Vanzi et al. \cite{vanzi08}). The ratio [FeII]($1.64\mu m$)/Br$\gamma$ can be used to distinguish thermal excitation in SN shocks from photoionization: in the first case the ratio is typically higher than a few tens, while in the second one it is significantly lower than 1, with typical values of 0.1 or less (Alonso-Herrero et al. \cite{almudina97}).

In He~2-10 the bulk of the [FeII] emission is almost coincident with the Br$\gamma$ one, mainly corresponding to the radio knots 4 and 1,2 of KJ99. However, the [FeII] peak is located $0.4\arcsec$ to the southwest of the Br$\gamma$ one, and the emission significantly extends toward the west, reaching the radio source knot 3 of KJ99. A fainter source of [FeII] is also visible beyond the north edge of the radio source 4. At the position of knots 1-2 the iron is faint, while at the location of knot 5 only a diffuse component can be seen.  The region with much fainter Br$\gamma$ emission around knots 1, 3, and 4 (see Sect. \ref{ionH}) is evident also in [FeII].

In NGC 5253 the [FeII] line is brightest at the location of cluster 5. It is interesting that we detect a second bright source $1.1\arcsec$ to the northwest of cluster 5, and another one $2.6\arcsec$ to the east, which have no obvious counterpart in the continuum or in Br$\gamma$.
\begin{figure*}
\centering
\includegraphics[angle=0, width=0.75\textwidth]{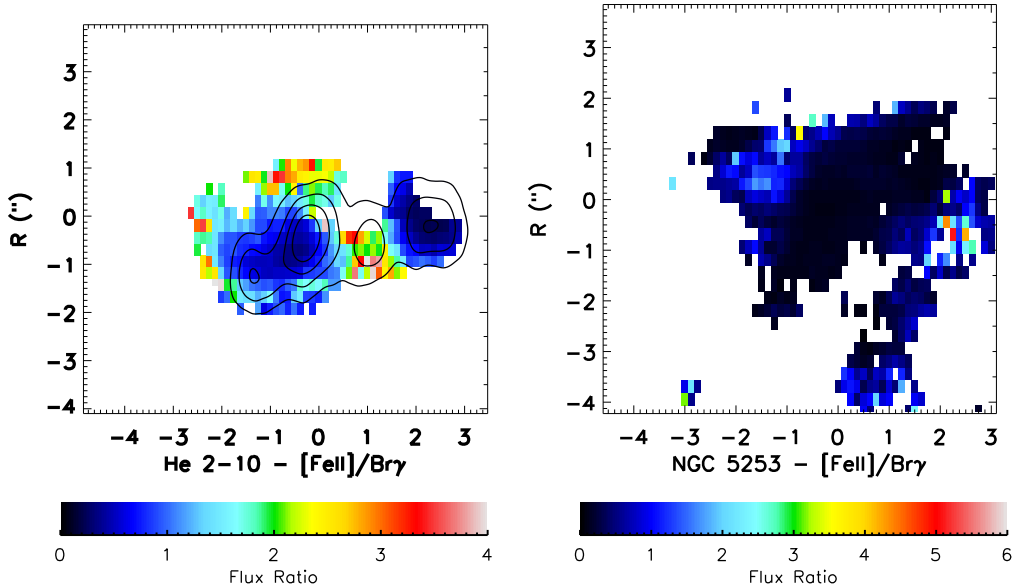}
\caption{Map of the [FeII]/Br$\gamma$ ratio for He~2-10 (left panel) and for NGC~5253 (right panel). The radio map for He 2-10 obtained with VLA 3.6 cm continuum image by Johnson \& Kobulnicky (\cite{johnson03}) is overplotted. The radio knot 3 classified as a possible non-thermal radio source by Johnson \& Kobulnicky (\cite{johnson03}) and the radio knot 4 classified by Cabanac et al. (\cite{cabanac05}) as a mixture of normal HII and SNR corresponds with the two regions of enhanced ratio, thus confirming the presence of SNR driven shocks.}
\label{FeII_Bry}
\end{figure*}

To better quantify the nature of these [FeII] sources, we used the ratio [FeII]/Br$\gamma$ as a diagnostic of the excitation mechanism. To minimize the effect of extinction and band shift we used the observed ratio [FeII](1.644)/Br12(1.641), and rescaled the Br12 flux to the corresponding Br$\gamma$ assuming an intrinsic ratio Br12/Br$\gamma$=0.189 (Hummer \& Storey \cite{hummer87}, for a case B recombination with $N=10^4$ cm$^{-3}$ and $T_e=10^4$ K). The maps of the obtained ratio are shown in Fig. \ref{FeII_Bry}. The highest values are in the range between 1 and 7, therefore consistent with a significant contribution by SNe in the regions with bright [FeII] emission. The data also support the suggestion of Johnson \& Kobulnicky (\cite{johnson03}) that knot 3 in He~2-10 is a non-thermal radio source. Indeed, this region (shown in the radio 3.6 cm continuum contours from Johnson \& Kobulnicky (\cite{johnson03}) in Fig. \ref{FeII_Bry}) is coincident with one of the regions with high [FeII]/Br$\gamma$ ratio in our data. Moreover, C05 claim that the more extended radio knot 4 is a complex mix of normal HII regions and SNR, because it shows a non-thermal signature with a slightly negative spectral index $\alpha$. Indeed we find in the northern region of this radio knot the other region with a significantly enhanced ratio, which confirms their analysis.
At the location of the radio knot 4 and 1,2 in He~2-10 and cluster 5 in NGC~5253 the [FeII]/Br$\gamma$ ratio is lower, with values in the range $\sim0.1-0.3$, indicating regions dominated by photoionization by young massive stars.
That in both galaxies the ratio FeII/Br$\gamma$ does not reach the maximum values observed in pure SN remnants seems to indicate that in all cases we do not observe pure shock-excited [FeII] lines, but possibly an additional contribution from the underlying stellar population. Despite this caveat, assuming that all the iron emission with [FeII]/Br$\gamma$$>1$ is dominated by the emission by SNR, we can estimate a SN rate for the two galaxies with the following relation:
\begin{equation} \label{SNr}
	SN_r (SN/yr) = \frac{L_{[FeII]}(tot)}{L_{[FeII]}(SNR)} \frac{1}{\tau(yr)} = A \times L_{[FeII]}(tot)
\end{equation}
where $L_{[FeII]}(tot)$ is the observed [FeII] flux with ratio [FeII]/Br$\gamma>1$ and a nignal/noise$>3$, $L_{[FeII]}(SNR)$ is the typical [FeII] emission from a SNR and $\tau(yr)$ is the mean life of the [FeII] emission from the SNR. Moorwood \& Oliva (\cite{moorwood88}), Vanzi et al. (\cite{vanzi97}), Colina (\cite{colina93}) and Mouri et al. (\cite{mouri00}) have computed the proportionality constant $A$ using observations of nearby galactic SNR, finding consistent results with an average $A=0.24\ (10^{40}\ erg/s)^{-1} yr$. Alonso-Herrero et al. (\cite{almudina03}) have instead computed the constant directly measuring the SN rate (SNr) in M82 and NGC~253 from radio Supernovae counts and the total [FeII] luminosity from HST narrow-band imaging. They find a lower $A= 0.08$, which may be due to a contribution from photo-excited regions to the total [FeII] luminosity. This value is close to the $A=0.05$ derived by van der Werf et al. (\cite{vanderwerf93}), estimating a SN rate from the radio synchrotron emission from the nuclei of NGC~6240 and measuring the [FeII] flux from the same regions. In this case the lower value is probably due to the different method to derive the SN rate. We therefore assume $A=0.24$, keeping in mind that the uncertainty may be at least a factor of three.  
Using eq. (\ref{SNr}) and measuring the [FeII] luminosity for the two galaxies in the pixels with $S/N>3$ and [FeII]/Br$\gamma>1$, we find:
\begin{equation}
	SNr_{(He~2-10)} = A \times 3.31\ 10^{38} erg/s = 7.9\ 10^{-3} SN/yr 
\end{equation}
\begin{equation}
	SNr_{(NGC~5253)} = A \times 9.68\ 10^{37} erg/s = 2.3\ 10^{-4} SN/yr
\end{equation}
We compared these values with the expected number of SNe as derived from the far-infrared (FIR) luminosity  observed in these galaxies, according to the relation between FIR and SN rate calibrated by Mannucci et al. (\cite{mannucci03}). With this method, we obtain a SN rate of $1.2 \times 10^{-2}$ and $1.1\times 10^{-3}$ SN/yr for He~2-10 and NGC~5253 respectively. The SN rates expected from the FIR luminosity are higher, especially for NGC~5253. However, we notice that our observations are not covering the whole extent of the activly star-forming regions of the galaxies, thus underestimating the total [FeII] flux. The result suggests a possible SNR origin for the observed [FeII] line emission in these galaxies. 

\subsection{Molecular hydrogen} \label{molH}

The H$_2$ line at 2.212 $\mu$m rest wavelength and additional fainter lines in the H and K bands allow us to trace the location and physical status of the warm molecular gas in the two galaxies. The molecular hydrogen lines in the NIR are actually mainly excited through fluorescence, i.e. absorption of UV photons, or thermally through collisions. The two different mechanisms mostly excite different ro-vibrational levels, so that it is possible to distinguish between the two processes by comparing the ratios of some of the lines with models (e.g. Black \& van Dishoeck \cite{black87}). The 2-1/1-0 S(1) ratio is however disgnostic in the low-density limit, as it always thermalizes at high density (see Sternberg \& Dalgarno \cite{amiel89}).

\begin{figure}
	\centering
	\includegraphics[angle=-90, width=0.50\textwidth]{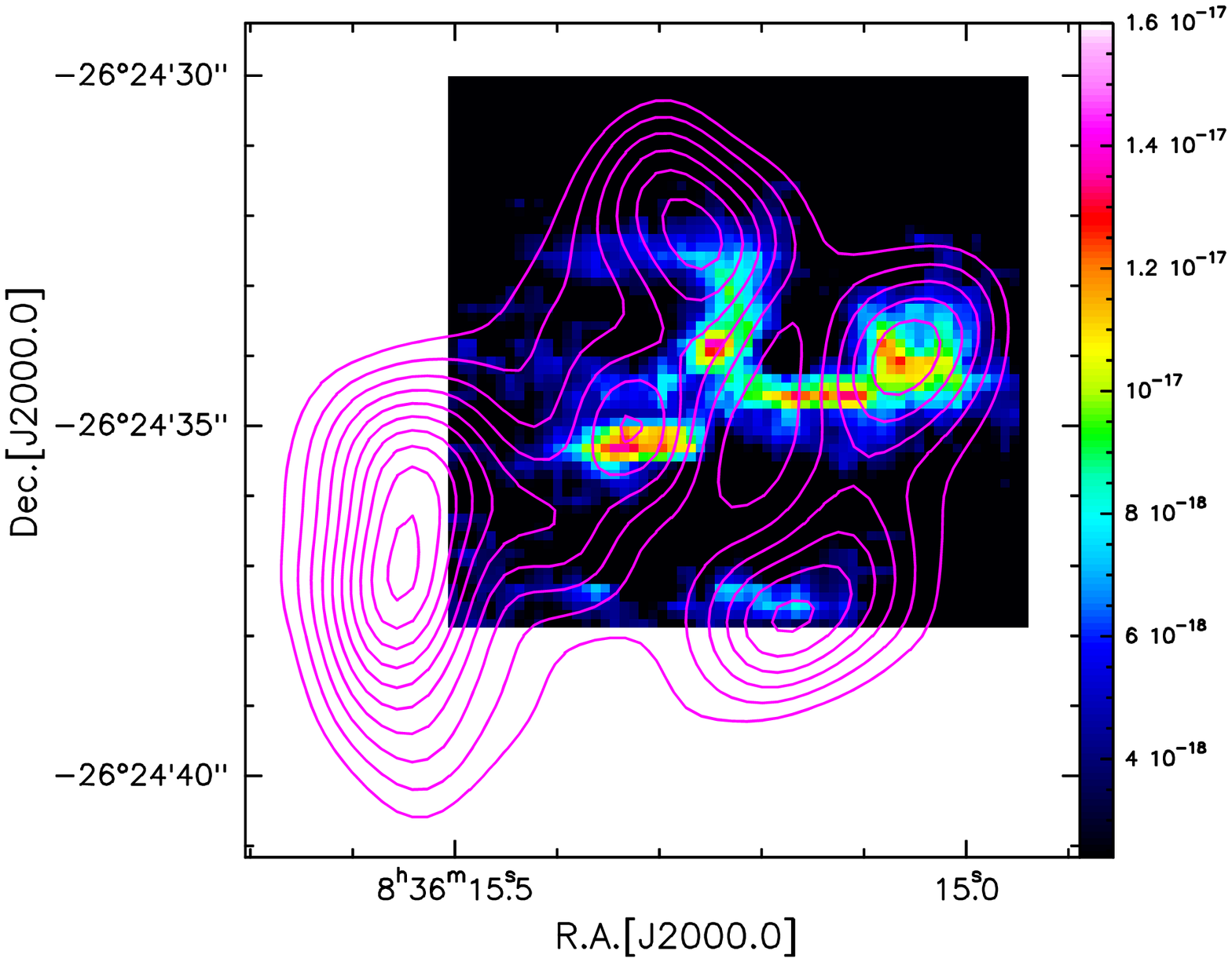}
	\caption{The H$_2$(2.212 $\mu m$) line emission map for He~2-10 compared with the CO(J=2-1) emission from Santangelo et al. (\cite{santangelo09}) shown in contours.}
	\label{ginacoflux}
\end{figure}
\begin{table*}
	\begin{center}
	\begin{tabular}{l l c c c c c c}
	\hline
	\hline
	\multicolumn{8}{c}{He~2-10} \\
	\hline
	\hline
	Line & $\lambda$ ($\mu m$) & East & Center & Edge & West & thermal & fluoresc. \\
	\hline
	2-1 S(3) & 2.073 & 0.22 & 0.21 & 0.18 & 0.28 & 0.08 & 0.35 \\
	2-1 S(2) & 2.154 & 0.08 & 0.09 & 0.07 & 0.12 & 0.04 & 0.28 \\
	2-1 S(1) & 2.248 & 0.31 & 0.18 & 0.35 & 0.27 & 0.08 & 0.56 \\
	\hline
	\hline
	\multicolumn{8}{c}{NGC5253} \\
	\hline
	Line & $\lambda$ ($\mu m$) & Cluster 5 & Cluster 5a & SW & W & thermal & fluoresc. \\
	\hline
	2-1 S(3) & 2.073 & 0.27 & 0.34 & 0.27 & 0.25 & 0.08 & 0.35 \\
	2-1 S(2) & 2.154 & 0.22 & $<0.13$ & 0.22 & $<0.09$ & 0.04 & 0.28 \\
	2-1 S(1) & 2.248 & 0.37 & 0.36 & 0.43 & 0.34 & 0.08 & 0.56 \\
	\hline
	\hline
	\end{tabular}
	\caption{Comparison of the observed ratios of H$_2$ lines to H$_2$(1-0) S1 at 2.12 $\mu m$ in four different regions of both galaxies (see text) with fluorescent and thermal excitation models by Black \& van Dishoeck (\cite{black87}). The uncertainties in the ratio are on the order of 30\% in He~2-10 and 15\% in NGC~5253.}
	\label{h2tab}
	\end{center}
\end{table*}

As shown in Fig. \ref{flux_he2-10}, in He~2-10 we detect three main regions of H$_2$ emission at 2.212 $\mu$m. All of them are extended, and appear to be diffuse with filamentary components. They are generally not clearly associated to any feature observed in the continuum or in other lines, except for the cloud corresponding to the radio knots 1 and 2, where Br$\gamma$ and continuum emission is also detected at corresponding location. In particular, the K band nucleus does not show a clear counterpart in H$_2$, and the H$_2$ filaments extends beyond the bright Br$\gamma$ regions, similar to what was observed by Vanzi et al. (\cite{vanzi08}) in II~Zw~40. 
However, if we compare our H$_2$ observations with the study of the molecular gas in this galaxy as traced by the CO(J=2-1) transition observed by Santangelo et al. (\cite{santangelo09}) in the millimeter at the Submillimeter Array (SMA) telescope, we find a strong spatial correlation between the CO and H$_2$ clouds. Although their data cover a larger field of view with lower angular resolution ($1.9\arcsec \times 1.3\arcsec$), the correlation between the positions of the clouds in the different species is clearly visible in Fig. \ref{ginacoflux}, with all the detected CO clouds showing a counterpart in the H$_2$. The correspondence in spatial distribution is not completely systematic, because the E-W structure to the SE of L1 and the peak just east of L4 do not have an obvious counterpart in the CO map. However, a good agreement is found also in velocity, as shown in Sect. \ref{kinematics} and Fig. \ref{ginacovel} (see also Henry et al. \cite{henry07}), which provides further support for the interpretation that the CO and H$_2$ we detect come from the same molecular structures.  The particular shape of the H$_2$ emission is probably because we preferentially see H$_2$ on the surface of the clouds, where it can be heated by UV photons or hit by shock waves.

The situation is different in NGC~5253 (see Fig. \ref{flux_ngc5253}), where the brightest peaks of the H$_2$ emission can be associated to clusters 5, 4, and 5a. But some differences in the spatial extend of the emission in the ionized and molecular hydrogen are present, because the region to the north-west of cluster 5 which emits in Br$\gamma$ has no counterpart of bright H$_2$ emission, and cluster 5 itself is elongated toward the southwest in H$_2$ in a region with little Br$\gamma$ and higher dust extinction (see Sect. \ref{estinzione}). 

To better understand the different excitation mechanisms responsible for the H$_2$ emission, we extracted 1D K-band spectra from different regions in the two galaxies, and compared the ratios of the detected H$_2$ lines to the 1-0 S(1) line at 2.12 $\mu m$. For He~2-10 we extracted one spectrum at the location of both the eastern and western $H_2$ clouds (with an aperture of $1.1\arcsec \times 0.5\arcsec$), as well as one spectrum at the emission peak at the center ($0.5\arcsec \times 0.5\arcsec$) and one in the gas bridge corresponding to the radio knot 3 ($1.1\arcsec \times 0.25\arcsec$). For NGC~5253 we extracted one spectrum at the peak in cluster 5 ($0.5\arcsec \times 0.5\arcsec$), one in the cluster 5a to the NE ($0.37\arcsec \times 0.5\arcsec$), and two more in the extended H$_2$ emission to the SW ($0.37\arcsec \times 0.5\arcsec$) and W ($0.87\arcsec \times 0.25\arcsec$) from cluster 5. The center of all regions is marked with a cross in Figs. \ref{flux_he2-10} and \ref{flux_ngc5253}. In Table \ref{h2tab} we compare the measured ratios with the predictions of the Black \& van Dishoeck (\cite{black87}) model for different excitation mechanisms. 

The 2-1S lines in the K band spectrum are particularly useful to distinguish between fluorescence and thermal excitation, because they are expected to be very weak in the latter case. Most of the lines were detected in different regions of both galaxies, which strongly suggests a dominant fluorescence excitation process. The measured ratios are lower than what is expected for pure fluorescence, allowing for some thermal contribution and suggesting that dense photodissociation regions are the dominant source of H$_2$ emission (see also Hanson et al. \cite{hanson02}). Although the S/N ratio for these faint lines is not optimal for a full 2D spatial analysis, we obtained maps of the most significant line ratios and verified that there is no relevant deviation from the behavior sampled by the ratios reported in Table \ref{h2tab}. In particular, we do not detect any region where shocks alone may account for the excitation of H$_2$, suggesting that dense photodissociation regions are the dominant source of H$_2$ emission. 

It is an interesting point that even in the regions where [FeII] is strong, the $H_2$ lines ratios does not support a dominant shock origin. This can possibly be understood by inspecting Fig.~\ref{FeII_Bry}, where we see that a large fraction of the [FeII]/Br$\gamma$ map shows values that are not strongly characteristic of shock-excited [FeII] emission (i.e. ratios in the range 1-2). Thus it appears that even when [FeII] emission is strong, shock-excitation mechanisms are never absolutely dominant, although contributing to the collected emission. The same situation is revealed by the $H_2$ lines ratios. Considering the actual nature of the regions where we observe the emission, i.e. galaxy centers populated by massive young clusters and dense HII regions, it is not surprising that this mixed situation is prevalent: there are copious amounts of UV light available as well as strong winds from stars and supernovae sweeping the ISM. We should expect to see both of these components contributing to the excitation of emitting sources.

\subsection{Dust} \label{estinzione}

\begin{figure}
	\centering
	\includegraphics[angle=0, width=0.50\textwidth]{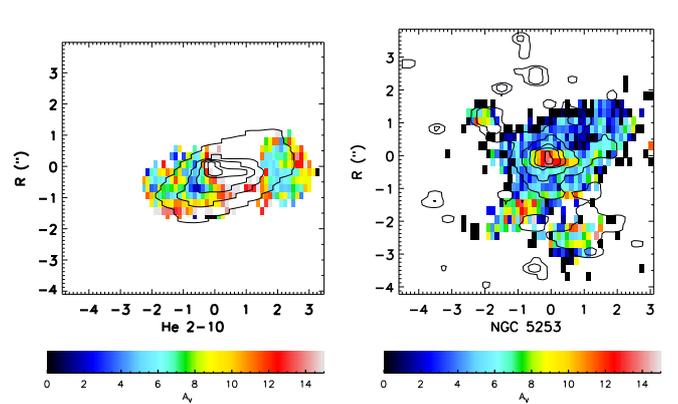}
	\caption{Extinction map in He~2-10 and NGC~5253 from Br12/Br$\gamma$ line ratio. The K band continuum from SINFONI is overplotted in contours for reference.}
	\label{dust}
\end{figure}

We obtained an extinction map for the two galaxies using the decrement of the observed ratio Br12/Br$\gamma$ with respect to the theoretical value of 0.189 calculated by Hummer \& Storey (\cite{hummer87}) for case B recombination, T=10$^4$ K and n=10$^4$ cm$^{-3}$, assuming the Rieke \& Lebowsky (\cite{rieke85}) extinction law. The resulting extinction maps are shown in Fig. \ref{dust} for the two galaxies, where the contours of the K continuum are overplotted for reference. 

In He~2-10 the extinction is ranging between $A_V\sim1.5-12$, with higher values at the edges of the knots, corresponding to the position of cluster L4a, and at the location of cluster L5 and L1.  C05 use the data of Vanzi \& Rieke \cite{vanzi97} to derive $A_V=10.5$ from Br$\gamma$/Br10, consistent with the values found here. This value is also consistent  with the $A_{Br\gamma}$=1.2 of Henry et al. (\cite{henry07}), corresponding to $A_V=10.6$. 

In NGC~5253 the extinction is higher at the location of the clusters, reaching its maximum value up to $A_V=12$ in cluster 5. This agrees with the results of Vanzi \& Sauvage (\cite{vanzi04}), who find that the brightest optical clusters of this galaxy are all characterized by a near-infrared excess that is explained by the combined effect of extinction and emission by dust. They estimate $A_V\sim7$ for cluster 5, but Calzetti et al. (\cite{calzetti97}) concluded that the cluster has to be embedded in a thick gas cloud with $A_V>9$ using the H$\alpha$/H$\beta$ ratio, in agreement with our estimate. 
Moreover, the dust reddening is less homogeneous than in He~2-10, and one of the dust lanes crossing the galaxy (see Cresci et al. \cite{cresci05}) is also evident in our  extinction map, south of cluster 5 in the NW-SE direction between cluster 5 and cluster 4.

\subsection{Stellar features} \label{coabsorption}

The strongest stellar absorption features in the available spectra are the CO bands at $\sim2.3\ \mu m$, whose depth increases with decreasing stellar temperature and increasing stellar luminosity, which is therefore most evident in regions with red giant and supergiant stars. Following Leitherer et al. (\cite{starburst99}), we use the spectroscopic CO index $CO_{sp}$ (Doyon et al. \cite{doyon94}, Puxley et al. \cite{puxley97}) and the equivalent width (EW) of the band to measure the age of the stellar population for the clusters old enough to show absorption features in their spectra. 

In He~2-10 the most prominent sources in absorption are the cluster L4 and the arc of old clusters detected only in continuum L3/6 (see Fig.~\ref{flux_he2-10}). We measure a $CO_{sp}=0.24$ and $CO_{sp}=0.22$ for the cluster and the arc, extracting the spectra in a circular aperture of 2 pixels and $4 \times 2$ pixels respectively. We also measure an EW in the rest frame wavelength range $2.2931\leq \lambda \leq 2.2983$ for the two regions of $16 \AA$ and $12 \AA$. Comparing these values with the stellar population synthesis model Starburst99 (Leitherer et al. \cite{starburst99}), and using the Br$\gamma$ detection to remove the degeneracy between young and old ages for a given $CO_{sp}$ or EW, we derive an age of $\sim 8.3$ Myr for the cluster L1 and an age between $14.5-40$ Myr for the clusters in the arc. 
%
%
These older clusters are lying at the center of the cavity between knot 1,3 and 4 detected in all the species discussed before (see Fig. \ref{flux_he2-10} and sections \ref{ionH}, \ref{iron} and \ref{molH}). We will discuss the properties of the cavity and its possible origin in greter detail in Sect.~\ref{cavity}.

In NGC~5253 we detected CO bands at low S/N only for a small cluster north of cluster 5, pointing toward a very young stellar population in this galaxy. 

\subsection{Gas kinematics} \label{kinematics}

\begin{figure}
	\centering
	\includegraphics[angle=0, width=0.50\textwidth]{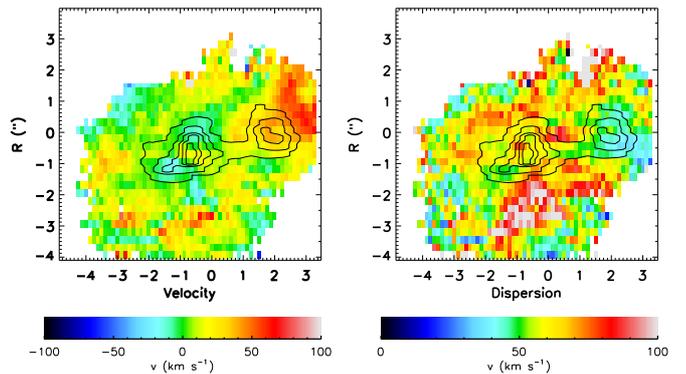}
	\caption{Gas kinematics from Br$\gamma$ line emission in He~2-10. Left panel: Br$\gamma$ emitting gas velocity. The velocity at the location of the brightest cluster is set to 0. Right panel: $\sigma$ velocity dispersion, corrected for instrumental effects. The contours of the Br$\gamma$ line emission are shown for comparison.} 
	\label{velhe210}
\end{figure}

\begin{figure}
	\centering
	\includegraphics[angle=0, width=0.50\textwidth]{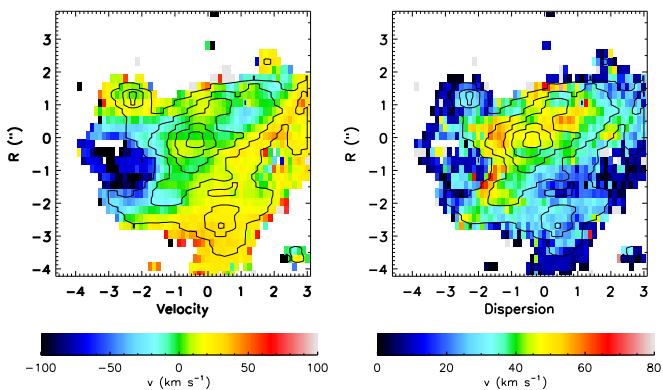}
	\caption{Gas kinematics from Br$\gamma$ line emission in NGC~5253. Left panel: Br$\gamma$ emitting gas velocity. The velocity at the location of the brightest cluster is set to 0. Right panel: $\sigma$ velocity dispersion, corrected for instrumental effects. The contours of the Br$\gamma$ line emission are shown for comparison.} 
	\label{velngc5253}
\end{figure}

We derived radial velocities and velocity dispersions by fitting a function to the continuum-subtracted spectral profile of the emission lines at each spatial position in the datacube for both galaxies. The function fitted was a convolution of a Gaussian with one spectrally unresolved emission line profile of a suitable sky line, i.e. a high signal and non-blended line in the appropriate band. 
A minimization was performed in which the parameters of the Gaussian were adjusted until the convolved profile best matched the data. During the minimization, pixels in the data that consistently deviated more than $3 \sigma$ from the average were rejected, and were not used in the analysis. The obtained Br$\gamma$ velocity  and velocity dispersion maps, corrected for instrumental broadening, are shown in Figs. \ref{velhe210} and \ref{velngc5253} for the two galaxies. The kinematics derived with the same method, but using the other emission lines discussed, are consistent with Br$\gamma$, but with lower S/N. The errors on the radial velocities range from 5 to 20 km/s, and areas with larger errors are not shown in the maps.  

\begin{figure}
	\centering
	\includegraphics[angle=-90, width=0.50\textwidth]{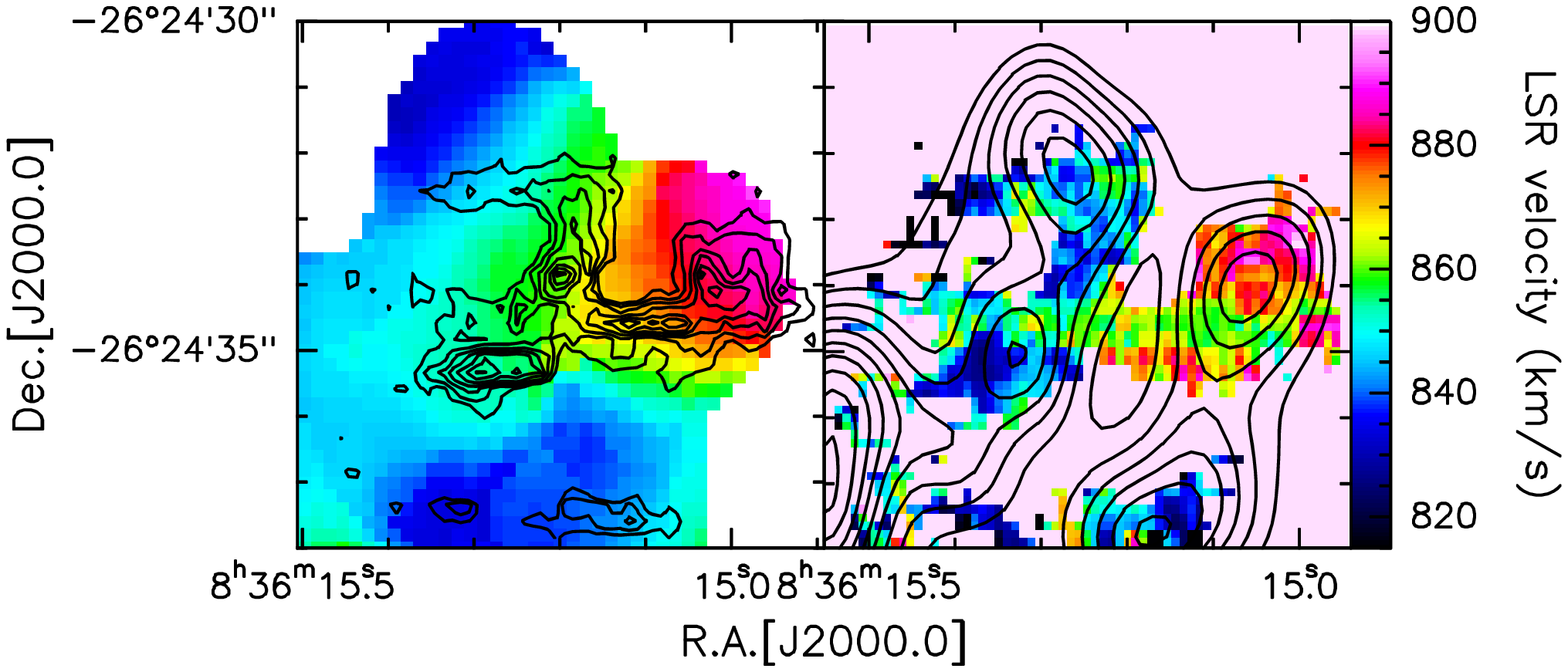}
	\caption{Comparison between the CO(J=2-1) and H$_2$ velocity field. Right panel: the CO(J=2-1) velocity map obtained by Santangelo et al. (\cite{santangelo09}) is shown, with the H$_2$ contours superimposed for reference. Left panel: the H$_2$(2.212 $\mu m$) velocity map from SINFONI with the CO(J=2-1) contours. The LSR velocity of the clouds detected in the two species agree well, with the eastern cloud showing a difference of $\sim35$ km/s with respect to the rest of the galaxy.} 
	\label{ginacovel}
\end{figure}

The gas kinematics is complex in both galaxies, with measured dispersions as high as $30-80$ km/s. The high measured line widths are probably produced by the combination of star formation driven turbulence and outflows in the medium. A coherent velocity pattern is also observed in the gas kinematics of the galaxies. In particular, a receding area is detected in the northwestern part of He 2-10, with a velocity difference of $\sim50$ km/s with respect to the continuum peak. This is fully consistent with the CO data of Santangelo et al (\cite{santangelo09}, see Sect. \ref{molH}). In Fig. \ref{ginacovel} we compare the CO(J=2-1) velocity map (left) and the H$_{2}$ one (right) on the same LSR velocity scale: the two gas components also have a similar velocity, with the western cloud showing a difference of $\sim 35$ km/s with respect to the rest of the galaxy. 

The molecular gas kinematics both for CO and $H_2$ is fully consistent with the one traced by the ionized gas. This supports a picture in which the HII regions move at the velocities of the molecular clouds from which they were born. A different result was obtained with similar observations by Vanzi et al. (\cite{vanzi08}) in the galaxy II Zw 40, where at least one giant molecular cloud is detected with different morphology and dynamics with respect to the HII regions.

We can derive a value for the $V/\sigma$ ratio for this galaxy using the observed velocity gradient and integrated dispersion, finding $V/\sigma\sim 1$. This value is lower than that observed in local spiral galaxies ($V/\sigma\gtrsim10$), but comparable with the low $V/\sigma$ ratios observed in high-z turbulent disks (see e.g. Cresci et al. \cite{cresci09}, Epinat et al. \cite{epinat09}, Law et al. \cite{law09}). Because the Jeans length $L_J$ in a gas rich system is proportional to $(\sigma/V)^2$ (Genzel et al. \cite{genzel08}), this supports the idea that the clumpy morphology of both high redshift disks and blue dwarf galaxies comes from gravitational instabilities in gas with high turbulent speed compared to the rotation speed, and a high mass fraction compared to the stars (see Elmegreen et al. \cite{elmegreen09}). 

In NGC 5253 a higher velocity dispersion area ($\sigma\sim50$ km/s) is centered around cluster 5 (see also Vanzi et al. \cite{vanzi06b}), probably produced by local episodes of winds or by turbulence induced by the stronger star-formation activity in the young central cluster. There are hints of an approaching area to the east of cluster 5, but the S/N is too low in that area to place firm constraints. No clear evidence of global rotation is detected in either galaxy in the area covered by our field of view. 

\section{Feedback-induced star formation in He~2-10?} \label{cavity}

We have already shown in Sect. \ref{ionH}, \ref{iron}, and \ref{molH} that a cavity depleted of gas is present in He~2-10 in all the species discussed, surrounded by some of the most actively star-forming regions of the galaxy, as knots 1, 3, and 4 (see Fig. \ref{flux_he2-10}). Interestingly, an arc of clusters with older stellar population is laying roughly at the center of this structure, as shown in section \ref{coabsorption} and in Fig.~\ref{flux_he2-10}. The [FeII]/Br$\gamma$ ratio, that is a tracer of shock excited gas, is measured to be higher at the edges of the cavity (see Sect. \ref{iron}). All these findings suggest a link between the formation of the cavity in the gas distribution and the triggering of the star formation activity in the surrounding regions, which is probably connected with the shocks responsible for the [FeII] emission. 

We can test this hypothesis by comparing the age of the clusters in the surrounding area with the expected travel time of a shockwave that may be responsible for the formation of the cavity in the gas. Given the large uncertainties on both the ages of the stellar populations and on the velocity of the shock, a more accurate comparison is not possible with the  available data.
Assuming that the center of the cavity is the oldest cluster in the area L3b (age$\sim 12.8$ Myr from H$\alpha$ Equivalent Width measured by C05), the main clusters in the region surrounding the cavity are L4a (distance $=0.7\arcsec$, age$\sim6.4$ Myr), L4b ($1.0\arcsec$, 5.8 Myr), and L1 ($1.21\arcsec$, 5.2 Myr). It can be already seen how there is a tendency for clusters closer to the center to be older, as expected if an expanding shell is the trigger for their formation. If we assume a shell velocity of $\sim 50$ km/s (e.g. Martin \cite{martin98}, Fujita et al. \cite{fujita03}), and a minimum time for SNe explosions in the central cluster of $\sim 6.3$ Myr (Leitherer et al. \cite{starburst99}) in order to create the shell through feedback, we estimate that the shock shell reached the location of the three clusters 5.9 Myr, 5.7 Myr and 5.5 Myr ago, respectively. These estimates agree with the measured ages of the clusters, given the large uncertainties, showing that the same event might be responsible for the formation of the cavity sweeping away the gas and triggering the star formation at its edge compressing the denser material.


\section{Conclusions} \label{fine}

We obtained integral field spectroscopy in the near-infrared of two nearby starburst galaxies: He~2-10 and NGC~5253. The study of the spectral features observed provided a detailed view of the environment where the star formation is occurring. We found that

   \begin{enumerate}
      \item the ionized gas traced by Br$\gamma$ is dominated by a single young super massive cluster in NGC~5253, while in He~2-10 the emission is due to several regions of active star-formation; 
      
      \item although the ISM is mostly photo-excited, as derived by the [FeII]/Br$\gamma$ and $H_2$ line ratios, some regions show a [FeII]/Br$\gamma$ excess that is probably due to a significant contribution by SN driven shocks. The location of the [FeII] bright regions corresponds in He~2-10 to the positions of radio sources classified as non-thermal by Johnson \& Kobulnicky (\cite{johnson03}) and Cabanac et al. (\cite{cabanac05}). We estimate the SN rates from the detected [FeII] emission, finding results that are compatible with the star-formation rates as derived by Br$\gamma$ emission and supporting this scenario; 
      
      \item the dust extinction in NGC5253 is higher at the location of the main young cluster 5, reaching values as high as $A_V=12$, shich shows that these young clusters are embedded in thick dust clouds; 
      
      \item the molecular gas clouds as traced by CO(2-1) and $H_2$ infrared lines in He~2-10 show consistent velocity and morphology, and an association with the formation of the youngest super star cluster as traced by the ionized gas; 
      
      \item we measured high turbulence in the ISM of both galaxies, $\sigma\sim30-80$ km/s, driven by the high star-formation activity. The clumpy morphology is related to the high gas fraction and high gas turbulence compared to the rotation speed, as in high-s disks.

      \item an arc of old clusters (age $\gtrsim 15-40$ Myr) is observed at the center of a cavity depleted of gas in He~2-10, surrounded by some of the most actively and young star-forming regions of the galaxy. This suggests a link between the feedback of the clusters in the arc, the formation of the gas cavity, and the triggering of the star formation at its edges. The estimates of the shock shell velocity agree with the measured ages of the young clusters ($\sim 5-12$ Myr);

\end{enumerate}

The obtained results confirm that the unique capabilities of integral field spectroscopy can provide unprecedented detail of the interstellar medium and star formation in nearby starburst galaxies, allowing us to study the different mechanisms that regulates and trigger the star formation in these extreme environments. 

\begin{acknowledgements}
PvdW would like to thank Liviu \c{S}tirb\u{a}\c{t} and Pascal Baars for discussions on these galaxies during the early stages of the project. This research has made use of the NASA/IPAC Extragalactic Database (NED), which is operated by the Jet Propulsion Laboratory, California Institute of Technology, under contract with the National Aeronautics and Space Administration. We received support from Basal Center for Astrophisics and Associated Technologies PFB-06. LV was supported by CONICYT through project Fondecyt n. 1095187. GC acknowledges PUC for the support received during his stay in Chile.
\end{acknowledgements}

\end{document}